\documentclass[aps,prx,twocolumn,superscriptaddress]{revtex4-2}
\usepackage{amssymb,bm,graphicx,amsmath}
\usepackage[dvipsnames]{xcolor}

\begin{document}
\author{I. Chestnov}
\email{igor\_chestnov@westlake.edu.cn}
\affiliation{Westlake University, School of Science, 18 Shilongshan Road, Hangzhou 310024, Zhejiang Province, China}
\affiliation{Westlake Institute for Advanced Study, Institute of Natural Sciences, 18 Shilongshan Road, Hangzhou 310024, Zhejiang Province, China}
\affiliation{Vladimir State University, Gorkii St. 87, 600000, Vladimir, Russia}

\author{Y. G. Rubo}
\affiliation{Instituto de Energ\'{\i}as Renovables, Universidad Nacional Aut\'onoma de M\'exico, Temixco, Morelos, 62580, Mexico}

\author{A. Nalitov}
\affiliation{Faculty of Science and Engineering, University of Wolverhampton, Wulfruna Street, Wolverhampton WV1 1LY, United Kingdom}
\affiliation{Institut Pascal, PHOTON-N2, Universit\'e Clermont Auvergne, 63001 Clermont-Ferrand, France}
\affiliation{ITMO University, St. Petersburg 197101, Russia}

\author{A. Kavokin}
\affiliation{Westlake University, School of Science, 18 Shilongshan Road, Hangzhou 310024, Zhejiang Province, China}
\affiliation{Westlake Institute for Advanced Study, Institute of Natural Sciences, 18 Shilongshan Road, Hangzhou 310024, Zhejiang Province, China}
\affiliation{Russian Quantum Center, Skolkovo, Moscow 143025, Russia}

\title{Pseudo-conservative dynamics of coupled polariton condensates}

\begin{abstract}
Open-dissipative systems obeying parity-time ($\mathcal{PT}$) symmetry are capable of demonstrating oscillatory dynamics akin to the conservative systems. In contrast to limit cycle solutions characteristic of nonlinear systems, the $\mathcal{PT}$-symmetric oscillations form a continuum of non-isolated orbits. 
However, precise sculpturing of the real potential and the gain-loss spatial profiles required for establishing of the $\mathcal{PT}$-symmetry is practically challenging. The optical devices, such as lasers, exhibit relaxation dynamics and do not operate as the $\mathcal{PT}$-symmetric systems. 
Here we demonstrate how these constraints can be overcome. 
We predict that a pair of optically trapped polariton condensates (a polariton dimer) can be excited and operated in the oscillating regime typical of the isolated systems. This regime can be realized in the presence of both dissipative and conservative coupling between the condensates and can be maintained at an arbitrary external pump intensity. Every orbit is characterised by a frequency comb appearing in the spectrum of a dimer in the presence of the conservative nonlinearity. Our results pave the way for the creation of the optical computing devices operating under the constant-wave external pumping.
\end{abstract}

\maketitle

\section{Introduction}

The general principles of quantum mechanics state that the Hamiltonian of an isolated system is a Hermitian operator. This way, the values of the energy form the entirely real spectrum and the corresponding eigenstates are linearly independent and mutually orthogonal. The real-valuedness of the energy means that the eigenstates are stationary and leave infinitely long, while their mutual independence implies that \textit{arbitrary} coherent superpositions of them are possible. 
These principles stand behind the idea of a quantum bit whose information capacity is manifested by ability of storing and processing such superposition states. The classical description of such systems leads to the concept of continuum of trajectories, which evolve conserving the phase volume. This corresponds to the conservation of unit cumulative probability in the quantum case.    

The situation changes dramatically if the system interacts with the surrounding environment. The interaction typically results in the energy dissipation and it leads to the destruction of the superposition states. At the mean-field level, this phenomenon can be captured by considering the total evolution operator as a non-Hermitian one. Its eigenenergies are complex, in general, and the stationary superposition solutions are impossible since the amplitude of at least one state diverges with time. 
For this reason, the open systems have been considered unsuitable for sustaining steady-state superpositions  until the pioneering work of Bender and  Boettcher, \cite{Bender1998} who opened the door to the room of pseudo-Hermitian physics.
They have shown that a certain class of open systems, whose Hamiltonians are invariant under combined parity $\hat{\mathcal{P}}$ and time inversion $\hat{\mathcal{T}}$ operations, have entirely real-valued energy spectra. 
Later, the theory of $\mathcal{PT}$-symmetric systems was developed and extended towards the more general concept of pseudo-Hermiticity \cite{Bender2002,Mostafazadeh2002}. The eigenstates of the $\mathcal{PT}$-symmetric Hamiltonian are not orthogonal but skewed \cite{ElGanainy2018}. Nevertheless these systems are capable of supporting superposition states as it was revealed for the first time in the oscillations dynamics of light in the coupled optical waveguides \cite{Ruter2010}.

For a $\mathcal{PT}$-symmetric system the Hamiltonian commutes with the joined $\mathcal{PT}$-operator, {$[\hat{\mathcal{H}},\hat{\mathcal{P}}\hat{\mathcal{T}}]=0$}. 
Since for a two-level system the time inversion corresponds to complex conjugation and the parity corresponds to the exchange of some pair of possible basis states, the $\mathcal{PT}$-symmetry implies the parity of Hermitian and the anti-parity of the anti-Hermitian part of $\hat{\mathcal{H}}$. Therefore, if one state dissipates, this dissipation must be balanced by the pumping of the other state with precisely the same rate. This exact balancing of the gain and losses appears to be a challenging task in practice. 

The principles of $\mathcal{PT}$-symmetry were successfully implemented in diverse physical settings including optomechanics \cite{Iorsh2020,Jing2014,Kepesidis2016}, cold atoms \cite{Zhang2016}, metamaterials \cite{Fleury2014}, etc. However, the most fertile platform  is optics, where balancing gain with losses can be easily realized by  sculpturing of the spatial distribution of  the complex refractive index \cite{Ruter2010}. In particular, the concept of $\mathcal{PT}$-symmetry was successfully applied to the light transmission problems revealing many non-trivial effects such as non-reciprocal propagation \cite{Chang2014,Ramezani2010}, unidirectional invisibility \cite{Lin2011}, loss-induced transparency \cite{Guo2009}. 

Another important class of applications is $\mathcal{PT}$-symmetric lasers where  the effect of spontaneous breaking of $\mathcal{PT}$-symmetry \cite{Ozdemir2019} was employed for assisting  single-mode lasing \cite{Hodaei2014,Feng2014,Peng2014,Gupta2019}.
In contrast to the transmission problems, in lasers the gain-loss balance is typically sacrificed to the domination of the gain which is a key condition of lasing. In the steady-state regime, the excess of the gain is compensated by the gain-saturation effect \cite{Chang2014,Gupta2019}. The latter is the nonlinear phenomenon and thus precludes $\mathcal{PT}$-symmetric systems  from the formation of the linear superposition states  \cite{Suchkov2016,Miroshnichenko2011,Konotop2016}. For these reasons, the gain-loss systems operating above the lasing threshold are typically devoid of pseudo-Hermitian dynamics: The simultaneous excitation of several eigenstates with  arbitrary proportions of their populations is forbidden.

In this paper we demonstrate that an engineering   of the gain-saturation mechanism is capable of endowing the gain-loss system with the behaviour typical to the isolated or open $\mathcal{PT}$-symmetric systems. In particular, we demonstrate the existence of a continuum of periodic solutions corresponding to the superposition of the stationary states in the driven-dissipative system operating above the lasing threshold. 

As a model system, we consider the condensate of exciton polaritons excited in an optical trap. It represents a versatile example of the driven open system   characterized by the self-excitation threshold. Sculpturing of the pump profile allows for creating optical traps for polaritons which offer a high level of tunability of the condensate eigenstates. Another peculiarity of the polariton systems is connected with the coupling mechanism. The coupling between the distinct states of the polariton condensate is entirely complex in contrast to the $\mathcal{PT}$-symmetric optical systems, for which the coupling is purely real, typically. This generalization allows us to extend the results of our analysis to the general pseudo-Hermitian systems.

For the sake of clarity, we start with a comparison of the conservative and pseudo-Hermitian dissipative systems.

\section{Isolated vs open dimer}\label{SecII}
Let us consider an isolated system comprised of two mutually coupled subsystems $\psi_1$ and $\psi_2$ (the \textit{dimer}).   When the modes  are in resonance and the coupling $J$ is symmetric and real (Josephson), the Hamiltonian reads
\begin{equation}\label{H_con}
\hat{\mathcal{H}}_J = J\hat{\sigma}_x,
\end{equation}
where  $\hat{\sigma}_i$ with $i=\{x,y,z\}$ are the Pauli matrices and the energy of each mode is set to zero. In this case there are two supermodes: the symmetric 
$\bm{\psi}_s = C(1,1)^\intercal$ and the antisymmetric $\bm{\psi}_{a}=C(1,-1)^\intercal$ states, where $C$ is an arbitrary constant. 
In what follows, bold symbols denote vectors: $\bm{\psi}=(\psi_1,\psi_2)^\intercal$.  
As follows from the Shr\"odinger equation ($\hbar=1$) 
\begin{equation}\label{SE}
	i\partial_t\bm{\psi}=\hat{\mathcal{H}}_J\bm{\psi},
\end{equation}
there is also a continuum of superposition states 
\begin{equation}\label{SPS}
	\bm{\psi}=\bm{\psi}_s e^{-i\omega_s t} + e^{i\varphi}\bm{\psi}_{a}e^{-i\omega_{a} t},
\end{equation}
where $\omega_{s,a}=\pm J$ are the real eigenfrequencies of $\hat{\mathcal{H}}_J$ and $\varphi$ is an arbitrary phase shift. The occupations of the supermodes, given by the norms $\left\| \bm{\psi}_s \right\| $ and $\left\| \bm{\psi}_{a} \right\|$, can be taken in arbitrary proportions. These solutions are revealed by the beatings of the bare modes populations with the period $\pi/J$. 

\begin{figure}
	\includegraphics[width=\linewidth]{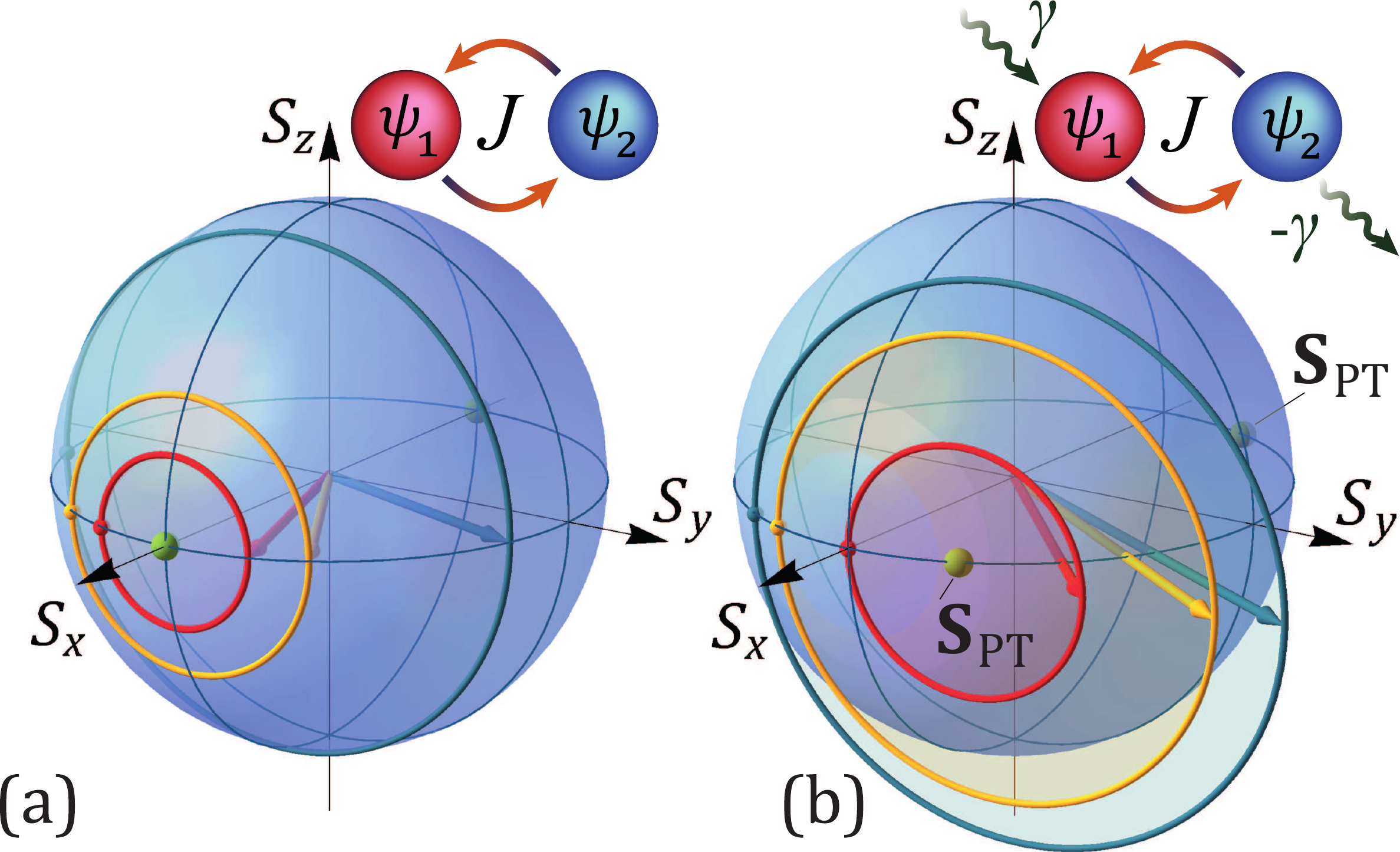}
	\caption{Dynamics of the isolated (a) and the open $\mathcal{PT}$-symmetric (b) dimers. Green dots indicate the Stokes vector positions at the stationary states. Colour lines are the trajectories corresponding to different starting points which are indicated by the dots of the same colour. In (b)  the ratio $\gamma/J=-0.4$ is used. }
	\label{fig1}
\end{figure}

A continuum of the oscillating states \eqref{SPS} also follows from the stability properties of the fixed point (FP) solutions $\bm{\psi}_s$ and $\bm{\psi}_a$. The relevant Jacobian matrix coincides with $\hat{\mathcal{H}}_J$. Therefore its eigenvalues are real as well and the FPs are centres \cite{Strogatz2018}. It implies that in the phase space, FPs are encircled with an infinite number of neutrally stable periodic trajectories. For the two-mode system, the phase space can be conveniently parameterized with three-dimensional Stokes vector $\mathbf{S}$ whose  components are 
$S_i=(\bm{\psi}^\dagger\hat{{\sigma}}_i\bm{\psi})$.
In this case the FPs locate at $S_x=\pm |\mathbf{S}|$, while the oscillating states draw circular trajectories around them, see Fig.~\ref{fig1}a.  

The prototypical example of the $\mathcal{PT}$-symmetric driven-dissipative dimer \cite{Ozdemir2019} is described by the Hamiltonian 
\begin{equation}\label{H_PT}
	\hat{\mathcal{H}}_{\rm PT}= \hat{\mathcal{H}}_J + i\hat{\mathcal{A}} 
	= J\hat{\sigma}_x + i \gamma \hat{\sigma_z},
\end{equation}
where 
$i\hat{\mathcal{A}}=i\gamma \sigma_z$ is the anti-Hermitian part stemming from the open nature of the system. 
The eigenenergies of \eqref{H_PT},
	$\lambda_{1,2} = \pm \sqrt{J^2-\gamma^2}$,
are real at $|J| > |\gamma|$ (in the opposite case the $\mathcal{PT}$-symmetry is broken spontaneously \cite{Guo2009,Ruter2010}). 
The  corresponding eigenstates $\bm{\psi}_{ss}^{\rm PT}=C\left(\pm e^{\pm i\xi},1\right)^\intercal$ with $\xi=\arcsin(\gamma/J)$ in the Stokes space locate at  
\begin{equation}\label{psi_PT}
\mathbf{S}_{\rm PT} = |C|^2\left(\pm \sqrt{1-\gamma^2/J^2},- \gamma/J,0\right)^\intercal.
\end{equation}
It is straightforward to demonstrate that for any Hamiltonian composed of the Hermitian and the anti-Hermitian parts akin to \eqref{H_PT}, the real energy spectrum implies a specific symmetry of the eigenvectors $\bm{\psi}_{ss}$. Namely, $\bm{\psi}_{ss}$ must obey $\langle\hat{\mathcal{A}}\rangle_{ss}=(\bm{\psi}^\dag_{ss}\hat{\mathcal{A}}\bm{\psi}_{ss})= 0$. For Hamiltonian \eqref{H_PT}, it yields a requirement of the symmetric occupation of the modes in the $\mathcal{PT}$-symmetric phase, $ S_z=0$. This is indeed the case for the eigenstates \eqref{psi_PT}. Besides, this condition can be considered as a $\mathcal{PT}$-symmetry criterion in the  systems governed by the more complex Hamiltonians including those which account for the nonlinear effects \cite{Konotop2016}. 

As long as the fixed points \eqref{psi_PT} are centres too, there is a continuum of superposition solutions analogues to \eqref{SPS}. The corresponding closed trajectories of the Stokes vector are shown in Fig.~\ref{fig1}b. Since these solutions are constructed from the symmetric states \eqref{psi_PT}, the trajectories are also symmetric with respect to the plane $S_z=0$. Therefore, the $\mathcal{PT}$-symmetry of the state is preserved on average: The cumulative expectation value of the anti-Hermitian operator $\hat{\mathcal{A}}$ along any closed trajectory $\mathcal{C}$ of the Stokes vector  vanishes, $\oint_\mathcal{C}(\bm{\psi}^\dag\hat{\mathcal{A}}\bm{\psi})d\bm{l}=0 $, where $d\bm{l}$ is an element along $\mathcal{C}$.   

Note that the existence of the oscillating solutions, which can be considered as a fingerprint of a superposition state, has been repeatedly observed in the driven-dissipative systems \cite{Leblanc2020,Nalitov2019,Ma2015,Kim2020,Rayanov2015,Marconi2020}. However, these oscillations are typically limit cycles (LC), whose trajectories in the phase space are isolated in a sense that any deviation from it relaxes back to the original periodic orbit. 


\section{$\mathcal{PT}$-symmetry of the polariton dimer}

In what follows we focus on the condensate of exciton polaritons created by a nonresonant laser field. The mean field dynamics of this system is described by the driven-dissipative Ginzburg-Landau equation for the condensate order parameter $\Psi(\mathbf{r})$.  The pump is tuned above the semiconductor band gap, which means that it can not gain the condensate state directly. Instead, it excites incoherent excitons which then relax their energy and momentum and scatter to the coherent condensate \cite{KavokinMicrocavities}. The rate of this scattering is governed by the total  density $n(\mathbf{r})$ of the reservoir which obeys the rate equation $\partial_t n = P – \left(\gamma_r + R|\Psi|^2\right)n$. Here $P$ is the pumping rate and $\gamma_r$ describes the reservoir decay. The presence of the feedback term $R|\Psi|^2n$ leads to the depletion of reservoir at large polariton densities. This constitutes the presence of the gain saturation effect for the coherent polaritons.

The reservoir evolution can be traced out in the adiabatic limit, which implies that the reservoir dynamics is fast as compared to the condensate evolution \cite{Keeling2008}. In this case the pump-saturation effect can be captured in the density-dependence of the condensate loss rate which corresponds to the appearance of the nonlinear dissipative terms in the Ginzburg-Landau equation, see Appendices~\ref{AppA} and \ref{AppB}.

We assume that polaritons are localized in a trap created by  
a spatially modulated pump $P(\mathbf{r})$ or the external potential $V_p(\mathbf{r})$. In a set of the trap eigenmodes $\varphi_i(\mathbf{r})$  we choose two isolated states, such that the whole phase space of the problem can be reduced to the dynamics of these states. The modes interact with each other when the spatial profiles of $P(\mathbf{r})$ or $V_p(\mathbf{r})$ are perturbed. This interaction provides the linear coupling required for establishing of the $\mathcal{PT}$-symmetric regime. 

Then in a two-mode approximation $\Psi(\mathbf{r}) = \psi_1(t)\varphi_1(\mathbf{r}) + \psi_2(t)\varphi_2(\mathbf{r})$ with $\int |\varphi_{1,2}|^2d\mathbf{r}=1$ after integrating out spatial degrees of freedom, the condensate dynamics is described by the following equation for the spinor $\bm{\psi}(t)$:
\begin{equation}\label{Pol_EQ}
	i\partial_t\bm{\psi} = \hat{\mathcal{H}}\bm{\psi} - i\nu \hat{N}_{\bm{\beta}} \bm{\psi} + \chi \hat{N}_{\bm{\alpha}} \bm{\psi},
\end{equation}
where the first term is a linear Hamiltonian
\begin{eqnarray}\label{H_pol}
	\hat{\mathcal{H}} &=& \hat{\mathcal{H}}_{\rm PH} +i\mathcal{W} \sigma_0 = \notag\\
		&=&\left(\begin{matrix}
		\varepsilon+i\gamma & J+i\varkappa \\
		J+i\varkappa & -\varepsilon-i\gamma
	\end{matrix}\right) +i\mathcal{W} \sigma_0 
\end{eqnarray}
characterized by the gain (loss) asymmetry parameter $\gamma$, the net gain  $\mathcal{W}$, the eigenenergy half-difference $\varepsilon$ and the real coupling parameters $J$ and $\varkappa$. 
It is crucial that the coupling between polariton condensates is entirely complex \cite{Aleiner2012}. This contrasts with the pure optical $\mathcal{PT}$-symmetric systems such as coupled waveguides or microcavities which are typically characterized by either real \cite{Ruter2010} or purely imaginary \cite{Suchkov2016,Alexeeva2014} coupling constants. The coexistence of the Josephson and dissipative couplings crucially affects the properties of polariton condensates in planar microcavities \cite{Aleiner2012,Nalitov2019,Ohadi2016}.

The last two terms in \eqref{Pol_EQ} contain the nonlinear operator
\begin{equation}\label{Eq_nonl_oper}
	\hat{N}_{\bm{\eta}} = \left(\begin{matrix}
		\eta_1\left|\psi_1\right|^2 + \left|\psi_2\right|^2 & 0\\
		0 & \eta_2 \left|\psi_2\right|^2 + \left|\psi_1\right|^2
	\end{matrix}\right),
\end{equation}
where $\bm{\eta}$ stands for $\bm{\alpha} = (\alpha_1,\alpha_2)^\intercal$ or $\bm{\beta}= (\beta_1,\beta_2)^\intercal$. The cross-phase modulation coefficients for both modes are set to 1, while the strength of the self-phase modulation effect is governed by the eigenmodes spatial structure, see  Appendix \ref{AppA}. 

The gain-saturation effect is governed by the non-Hermitian nonlinear term in Eq.~\eqref{Pol_EQ}. It stems from the depletion of the reservoir density above the threshold which balances the growth of the condensate population. The  Hermitian nonlinearity arising from the polariton-polariton interactions is described by the last term in Eq.~\eqref{Pol_EQ}. 

First, we neglect the nonlinear effects. The dissipative coupling $\varkappa$ obstructs $\mathcal{PT}$-symmetry of \eqref{H_pol} as long as its anti-Hermitian part is not anti-symmetric with respect to the mode permutation. 
Nevertheless, at the specific detuning, namely, 
\begin{eqnarray}\label{Eq_detuning}
	\varepsilon = -\varkappa J / \gamma,
\end{eqnarray}
the eigenenergies of \eqref{H_pol},
\begin{equation}\label{H_PH_EE}
	\lambda_{1,2} = i\mathcal{W} \pm \sqrt{J^2-\gamma^2}\sqrt{1+\varkappa^2/\gamma^2},
\end{equation}
are real provided that
\begin{subequations}\label{PH_conditions}
\begin{eqnarray}
\label{H_PH_cond_a}	\mathcal{W} &=& 0, \\
\label{H_PH_cond_c}	|J| &>& |\gamma|.
\end{eqnarray}
\end{subequations}
The second requirement coincides with the $\mathcal{PT}$-symmetry existence condition. When it holds, the condition \eqref{H_PH_cond_a} defines a threshold level of pumping below which $\Im({\lambda_{1,2}})<0$ and there is a single stationary state with $\mathbf{S}=0$. 

A real energy spectrum implies that at the threshold the Hamiltonian $\hat{\mathcal{H}}_{\rm PH} = \hat{\mathcal{H}}(\mathcal{W}=0)$ possesses pseudo-Hermitian properties. Although it does not commute with the $\mathcal{PT}$-operator, $\hat{\mathcal{H}}_{\rm PH}$ can be mapped to the conventional $\mathcal{PT}$-symmetric Hamiltonian \eqref{H_PT} with the use of the unitary rotation \cite{Bender2003,Wang2013}
\begin{equation}\label{Rotation}
	\hat{\mathcal{R}}=e^{i\hat{\sigma}_y\theta/2}=\left(\begin{matrix}
		\cos{\theta/2} & -\sin{\theta/2} \\
		\sin{\theta/2} & \cos{\theta/2}
	\end{matrix}\right), \ \tan{\theta}=\varkappa/\gamma,
\end{equation} 
which implies $\hat{\mathcal{H}}_{\rm PT}=\hat{\mathcal{R}}\hat{\mathcal{H}}_{\rm PH}\hat{\mathcal{R}}^{-1}$.
Therefore, we define the generalized Hermitian parity operator {$\hat{\tilde{\mathcal{P}}}=\hat{\mathcal{R}}\hat{\sigma}_x\hat{\mathcal{R}}^{-1} = \gamma \hat{\sigma}_x - \varkappa \hat{\sigma}_z$} for which { $\left[\hat{\mathcal{H}}_{\rm PH},\hat{\tilde{\mathcal{P}}}\hat{\mathcal{T}}\right]=0$}. Thus the Hamiltonian $\hat{\mathcal{H}}_{\rm PH}$ is $\tilde{\mathcal{P}}$-pseudo-Hermitian  \cite{Konotop2016,Wang2013}.

The Stokes vectors of the eigenstates of $\hat{\mathcal{H}}_{\rm PH}$ can be immediately obtained rotating \eqref{psi_PT} on the angle $\theta$ about the $S_y$ axis:
\begin{equation}\label{H_PH_eigst}
    \mathbf{S}_{\rm PH} = |C|^2 \left(\pm \frac{ \sqrt{1-\gamma^2/J^2}}{\sqrt{1+\varkappa^2/J^2}},- \gamma/J, \mp \frac{ \varkappa\sqrt{1-\gamma^2/J^2}}{\gamma\sqrt{1+\varkappa^2/J^2}}\right)^\intercal.
\end{equation}

Using the threshold pump as it is required by the condition  \eqref{H_PH_cond_a} is impractical because of the strong quantum fluctuations of the order parameter. That is why polariton lasers typically operate at $\mathcal{W}>0$. In this regime, the linear dynamics predicts an exponential growth of the total occupation. However, even in this case the $\mathcal{\tilde{P}T}$-symmetric dimer retains its underlying symmetry. Since the net gain $\mathcal{W}$ drives the dimer symmetrically, it affects the length but not the direction of the Stokes vector, which turns out to be the same as at $\mathcal{W}=0$. Therefore, it is convenient to resort to the unit state vector $\mathbf{s}=\mathbf{S}/|\mathbf{S}|$ for which the expectation value of the anti-Hermitian operator vanishes,
\begin{equation}\label{PH_criterion}
    \langle\hat{\mathcal{A}}_{\rm{PH}}\rangle=\gamma s_z + \varkappa s_x=0,
\end{equation} 
as it occurs for the stationary states \eqref{H_PH_eigst}. This phenomenon is known as a passive (or quasi-) $\mathcal{PT}$-symmetry \cite{Guo2009,Ornigotti2014} and can be revealed in any open system with the loss-rate imbalance between the subsystems \cite{Ozdemir2019}. 

\section{Pseudo-Hermitian dynamics of polariton condensate above the threshold}
\subsection{The nonlinear phase  diagram} \label{Sec:IV}
The saturation of the gain prevents an unbounded growth of the state norm at $\mathcal{W}>0$. Besides, it fixes the total occupation $S$ of the supermodes which  now is not arbitrary but dictated by the requirement of a balance between the saturated gain and dissipation. This constraint is expected to rule out the continuum of superposition states with variable supermode occupations. However, it is possible to demonstrate that the parity-time symmetric system is capable of supporting a continuum band of oscillating solutions provided that the gain saturation mechanism is properly engineered.  

In what follows it will be convenient to recast the nonlinear problem \eqref{Pol_EQ} in the Stokes vector basis:
\begin{equation}\label{Eq_pseudospin}
    \partial_t \mathbf{S} = \left[\mathbf{S} \times \mathbf{B} \right]+ \mathbf{E} S + \Gamma \mathbf{S},
\end{equation}
where 
\begin{subequations}\label{Eqs_BEG}
    \begin{eqnarray}
        \mathbf{B}&=&(2J,0,2\varepsilon + \chi\left[(\alpha_1-\alpha_2)S + (\alpha_1 + \alpha_2-2)S_z\right])^\intercal,\\
        \mathbf{E}&=&(2\varkappa,0,2\gamma - \nu \left[ (\beta_1-\beta_2)S+(\beta_1+\beta_2-2)S_z\right])^\intercal,\\
        \Gamma&=& 2 \mathcal{W} - \nu\left[(\beta_1+\beta_2+2)S +(\beta_1-\beta_2)S_z\right].
    \end{eqnarray}
\end{subequations}
From now on we assume that the $\mathcal{\tilde{P}T}$-symmetry of $\hat{\mathcal{H}}_{\rm{PH}}$ is preserved and focus on the impact of the nonlinear effects.

Equation~\eqref{Eq_pseudospin} has either two or four FP solutions. In addition to two trivial solutions $\rm{FP}_1$ and $\rm{FP}_2$ originating from the linear eigenstates \eqref{H_PH_eigst}, a couple of symmetry-breaking FPs appear at the strong pumping when the Hermitian nonlinearity dominates. These states are typically characterised by a large extent of the population imbalance $S_z/|\mathbf{S}|$.

Besides, the numerical analysis of Eq.~\eqref{Eq_pseudospin} reveals the oscillating solutions. Realization of the particular regime depends crucially on the strength and the structure of the nonlinear interactions. In what follows, we assume a symmetry between the self-phase modulation coefficients in Eq.~\eqref{Eq_nonl_oper}, $\eta_1=\eta_2$. Moreover, since the nonlinear dissipation arises from the condensate-density dependence of the reservoir occupation $n\propto |\Psi|^2$, this term is expected to have the same structure as the Hermitian nonlinearity. Thus we take  $\alpha_1=\alpha_2=\beta_1=\beta_2 = \eta$. 

\begin{figure}
    \includegraphics[width=\linewidth]{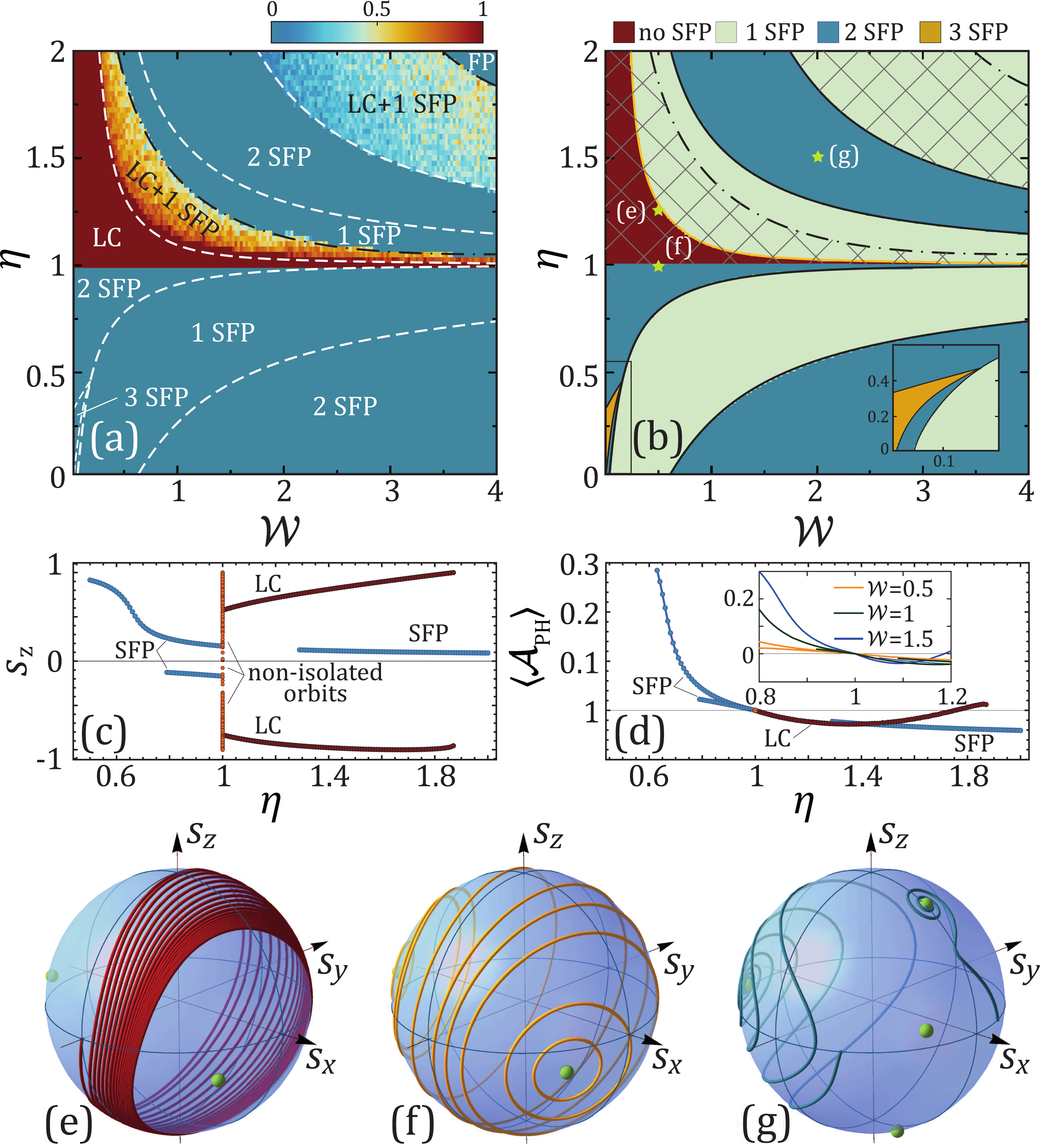}
    \caption{Dynamics of the nonlinear pseudo-Hermitian dimer. (a) A phase diagram illustrating the limit cycle (LC) formation probability, for $J=1$, $\varkappa=0.1$, $\chi=1$, $\nu=0.2$, $\gamma=0.5$, $\varepsilon$ obeys condition \eqref{Eq_detuning}. The dimer is excited starting from the noisy initial conditions. The data were averaged over 50 realizations for each point. (b) The chart of attractors on the $(\eta,\mathcal{W})$-plane. The false-color encodes the number of stable fixed-point (SFP) states of the dimer while the hatched domains contain a stable LC. The insert shows the magnified region from the lower-left corner. (c) The local extrema of the $s_z$-component plotted against $\eta$ at $\mathcal{W}=0.5$. The blue points correspond to SFPs, the vinous characterize the trajectory of the LC, the orange ones demonstrate multiple realizations of the neutrally stable orbits at $\eta=1$. (d) The expectation value of the anti-Hermitian operator $\hat{\mathcal{A}}_{\rm PH}$ averaged over the stable spin trajectory at $\mathcal{W}=0.5$. The inset shows the vicinity of $\eta=1$ for three different values of the net gain. (e) -- (f) The spin trajectories on the unit sphere for three different regimes indicated by the yellow stars in (b): the LC (e), multiple non-isolated orbits corresponding to different starting points (f), coexistence of two FP attractors reached from different initial conditions (g).}
    \label{Fig:PD}
\end{figure}

A phase diagram on the  $(\eta,\mathcal{W})$-plane is shown in Fig.~\ref{Fig:PD}a,b. Here we fix $J=1$ rescaling the time in Eq.~\eqref{Eq_pseudospin} and set $\chi=1$ which is equivalent to the normalization of the state amplitude. {A variation of the gain $\mathcal{W}$ with the given  nonlinear dissipative coefficient $\nu$  renders the strength of the nonlinear effects. Therefore, we keep  the value of $\nu$ fixed to $\nu=0.2$ without loss of generality.}  The structure of the phase diagram depends crucially on whether the Josephson and dissipative coupling parameters are of the same or of the opposite signs. Here we consider $J/\varkappa>0$, while the opposite case is described in the Appendix~\ref{AppNegJtDratio}.

The oscillating solutions exist at $\eta \geq 1$ as it is shown in Fig.~\ref{Fig:PD}a, which demonstrates a probability that the dimer resides on the periodic orbit starting to grow from a small initial seed with a random pseudospin orientation. There are two types of periodic solutions. The first one is realized at weak pumping $\mathcal{W}$ in the vicinity of $\eta=1$. Note that the domain of its existence is reflected across the line $\eta=1$ upon changing a sign of $J/\varkappa$.  For this state, the Stokes vector precesses around the trivial fixed points $\rm{FP}_{1,2}$ akin to the superposition states realized in the linear $\mathcal{PT}$-symmetric system, see Fig.~\ref{Fig:PD}e. This orbit however is isolated in the phase space indicating that it is a LC attractor.  The LC of the second type always exists at $\eta>1$. Its trajectory passes near the symmetry-breaking fixed point, see the Appendix~\ref{AppPhaseDiagramDetails}.

This analysis also demonstrates the presence of the large domains where several stable attractors coexist. Fig.~\ref{Fig:PD}b shows the regions of multistability between two FPs (see Fig.~\ref{Fig:PD}g for the example of the relevant dynamics), three FPs and between a single FP and a stable LC solution. The latter regime corresponds to the gradient colors in the probability map Fig.~\ref{Fig:PD}a.

The phase diagram indicates an exceptional role of the case of equal self- and cross-phase modulation coefficients, $\eta=1$. It corresponds to the switching between the FP and LC regimes. Besides, in its vicinity, the dynamics demonstrates a slowing down phenomenon when the time needed for a perturbed state to return to the stable attractor diverges. Figure~\ref{Fig:PD}c shows a bifurcation diagram for a fixed net gain and varying self-to-cross phase modulation ratio $\eta$. The diagram contains a collection of the local extrema of the normalized population imbalance $s_z=S_z/S$ for the attractors. The FP solutions are discontinuously transformed to the LC solution as $\eta$ exceeds 1. It turns our that exactly at $\eta=1$ there is a continuum of oscillating trajectories. The orange points in Fig.~\ref{Fig:PD}c correspond to multiple realisations of the periodic solutions starting from random initial conditions. The examples of the relevant trajectories on the surface of the unit sphere are shown in Fig.~\ref{Fig:PD}f.

The continuum of periodic orbits is a fingerprint of the pseudo-conservative dynamics. The parity-time symmetric properties of these states can be examined with the use of the $\mathcal{\tilde{P}T}$-symmetry criterion \eqref{PH_criterion}.  For all stable attractors, the expectation value of the anti-Hermitian part of the linear Hamiltonian $\langle\hat{\mathcal{A}}_{\rm{PH}}\rangle$  vanishes as $\eta$ approaches 1, see Fig.~\ref{Fig:PD}d. This result holds for any value of net gain parameter, as it is shown in the inset. Note that for the periodic solutions, $\langle\hat{\mathcal{A}}_{\rm{PH}}\rangle$ was be averaged over the spin-trajectory.

\subsection{Pseudo-conservative dynamics at the  equal self- and cross-phase modulation coefficients}

The surprising behaviour at the equal cross- and self-phase modulation coefficients can be attributed to the specific symmetry of the nonlinear term. The net gain $\mathcal{W}$ pumps the dimer symmetrically, which results in the passive $\mathcal{PT}$-symmetry in the unsaturated regime. The gain-saturation effect ruins the symmetry of the state if the relevant nonlinear term has a different symmetry. However, at $\eta=1$ the dissipative nonlinearity depends solely on the total number of particles $S$, see Eq.~\eqref{Eq_nonl_oper}, i.e. it  shares the same symmetry as the gain term. That is why it is not detrimental to the pseudo-Hermitian properties of the system.  The conservative nonlinearity in this case governs evolution of the global phase of the dimer but disappears from Eq.~\eqref{Eq_pseudospin}. As a result, the nonlinear terms enter Eq.~\eqref{Eq_pseudospin} only via the $\Gamma$ parameter which affects the length but not the orientation of the Stokes vector.

That it is why it is reasonable to recast Eq.~\eqref{Eq_pseudospin} with the use of the normalized  vector $\mathbf{s}$. At $\eta=1$ it yields
\begin{equation}\label{Eq_eff}
	\partial_t \mathbf{s} = \left[\mathbf{s} \times {\mathbf{b}}\right],
\end{equation}
where the components of the effective magnetic field $\mathbf{b}$  are \begin{subequations}\label{B_eff}
	\begin{eqnarray}
		b_x &=& -2(J + \gamma S_y), \\
		b_y &=& 2(\gamma S_x - \varkappa S_z),\\
	    b_z &=& -\frac{\varkappa}{\gamma} b_x.
	\end{eqnarray}
\end{subequations}
Since  the problem is devoid of both the net gain and the nonlinear parameters after the normalization, it has the same stationary solutions \eqref{H_PH_eigst} (with $|C|=1$) as the $\tilde{\mathcal{P}}\mathcal{T}$-symmetric Hamiltonian $\hat{\mathcal{H}}_{\rm PH}$.

\begin{figure}
	\includegraphics[width=0.66\linewidth]{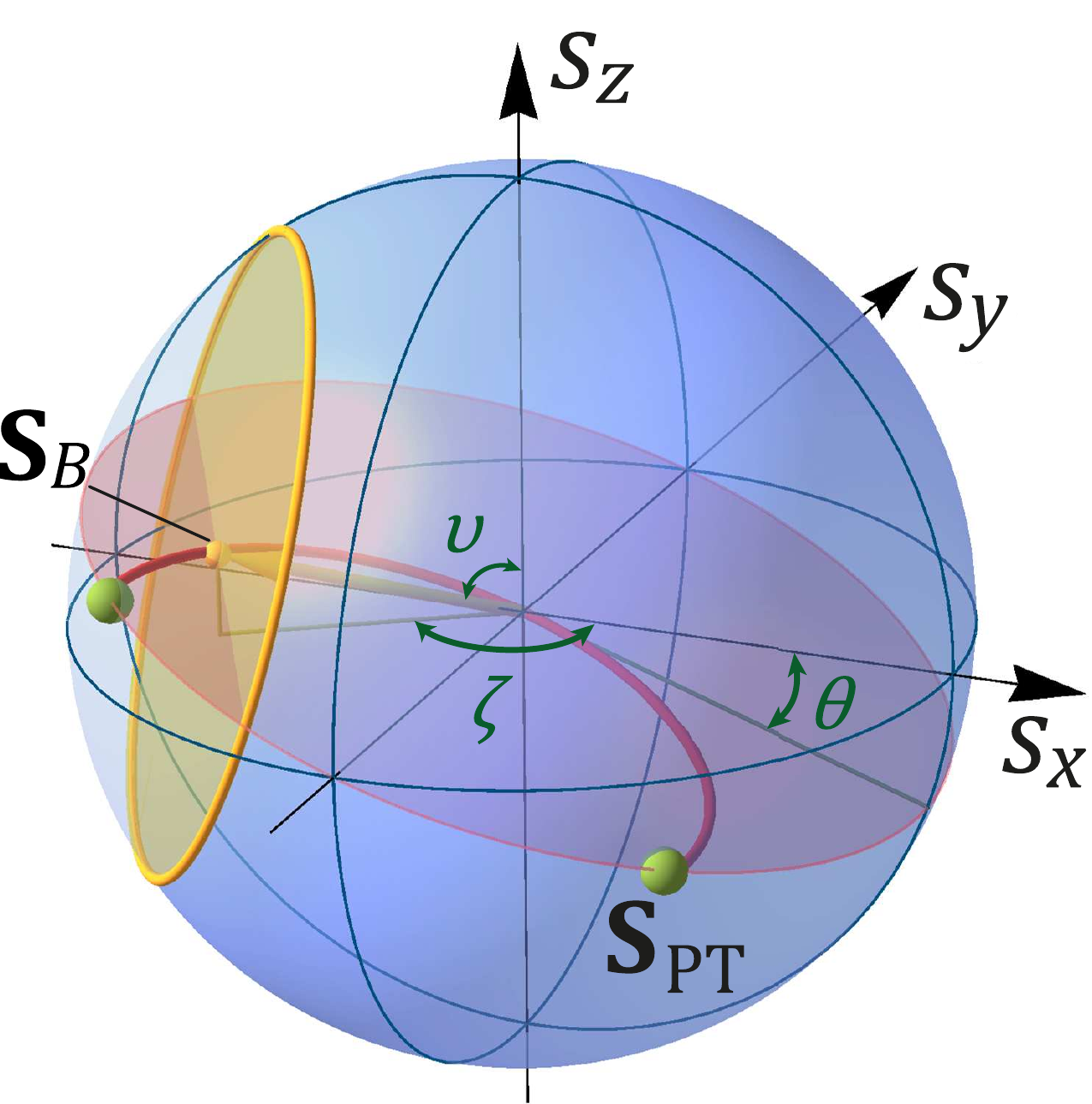}
	\caption{Pseudo-conservative dynamics above threshold. Precession of the Stokes vector about the effective magnetic field $\mathbf{b}$ whose direction coincides with the vector $\mathbf{S}_B$. The Stokes vector trajectory is shown by the yellow circle. The possible positions of the $\mathbf{S}_B$-vector are indicated by the red arc which is situated in the plane inclined by the angle $\theta$ with respect to the $(s_x,s_y)$ plane. The parameters are $J=1$, $\varkappa=0.25$, $\chi=1$, $\nu=0.2$, $\mathcal{W}=0.5$, $\eta=1$, $\gamma=0.75$, $\varepsilon$ obeys \eqref{Eq_detuning}.}
	\label{fig3}
\end{figure}

Besides, Eq.~\eqref{Eq_eff} predicts precession of the Stokes vector about the effective magnetic field. The precession frequency $|\mathbf{b}|$ is not constant because $b_i$ are governed by the instantaneous position of the Stokes vector. 
However, the direction of the vector ${\mathbf{b}}$ is conserved dynamically. It is straightforward to demonstrate that  the polar $\upsilon$ and the azimuthal $\zeta$ angles of $ {\mathbf{b}}$ (see Fig.~\ref{fig3})  are the integrals of motion of Eq.~\eqref{Eq_eff}.
The existence of the conserved quantities indicates that the dynamics  is \textit{conservative}. Therefore, the stationary solutions \eqref{H_PH_eigst} can not be attracting FPs \cite{Strogatz2018}. 

Since the orientation of $\mathbf{b}$ is fixed by the initial position of $\mathbf{s}$, the Stokes vector follows the circular trajectories on the Bloch sphere, see Fig.~\ref{fig3}. The positions of the centres of these circles are given by the vector 
\begin{equation}\label{S_B}
	\mathbf{S}_B= \mathbf{b} \left( \mathbf{b}\cdot \mathbf{s} \right)\left/ |\mathbf{b}|^2 \right..
\end{equation}

Obviously ${S}_B$ is an integral of motion too. With the use of \eqref{B_eff} and \eqref{S_B} we obtain a parametric equation which defines all possible positions of  $\mathbf{S}_B$:
\begin{equation}\label{Eq_circle}
	S_B^2 = -\frac{J}{\gamma} S_{By},
\end{equation}
where $S_{By}=S_B{b_y}\left/{\left|\mathbf{b}\right|}\right.$ stands for the projection of $\mathbf{S}_B$ on the $s_y$-axis. Equation~\eqref{Eq_circle} defines the circle in the plane turned on $\theta$ about the $s_y$-axis. The radius of this circle is $J/2\gamma$ while its centre is located at $(0,-J/2\gamma,0)$,  see the red arc in Fig.~\ref{fig3}. The circle crosses the point of origin, where $S_B=0$ and the Stokes vector follows the great circle of the Bloch sphere. Also, it intersects the surface of the Bloch sphere at the fixed points \eqref{H_PH_eigst}. In between, the Stokes vector precesses along a circular trajectory whose radius can be varied continuously from zero (at the FPs) to unity (at the origin) depending on its initial position. Thus the ensemble of possible trajectories of the Stokes vector spans the entire Bloch sphere. 

\begin{figure*}
	\includegraphics[width=0.75\linewidth]{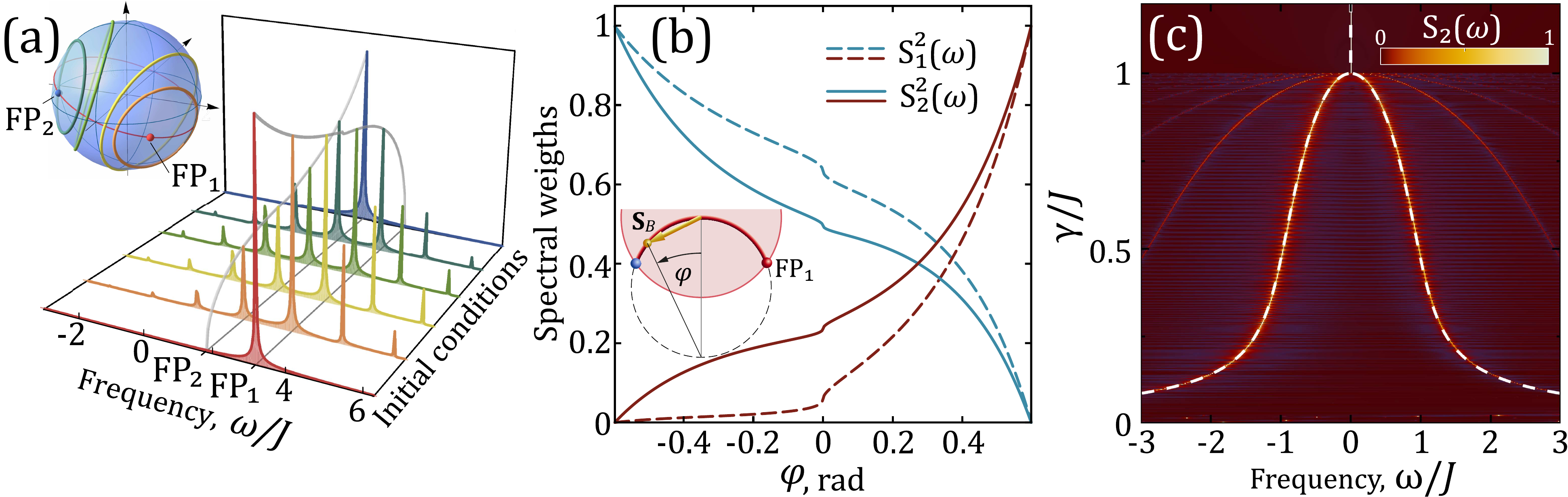}
	\caption{(a) The emission spectrum $S_2(\omega)$ for different initial positions of $\mathbf{S}_B$ vector. The corresponding trajectories of the Stokes vector are illustrated in the inset. The red and the blue curves correspond to  $\rm{FP}_1$ and $\rm{FP}_2$.  (b) The weights of the spectral components originating from the fixed points $\rm{FP}_1$ (solid lines) and $\rm{FP}_2$ (dashed lines) for the spectra shown in panel (a). An angle $\varphi$ on the horizontal axis parameterizes initial position of  $\mathbf{S}_B$ as it is shown in the inset. (c) The breaking of pseudo-Hermiticity in the linear regime $\chi=0$ triggered by the increase of the gain asymmetry $\gamma$. The remaining parameters are fixed except for the detuning $\varepsilon$ whose value is being tuned continuously in order to satisfy the $\tilde{\mathcal{P}}\mathcal{T}$-symmetry condition \eqref{Eq_detuning}.
	The parameters are the same as in Fig.~\ref{fig3}. For (a) and (b)  $\gamma=0.75$.}
	\label{fig4}
\end{figure*}

The presence of the pseudospin precession can be revealed in the condensate emission spectrum, see Fig.~\ref{fig4}. In the pseudo-conservative regime, the spectrum has a structure of a frequency comb. This contrast to the spectral shape of the linear superposition state \eqref{SPS}.  The formation of the frequency comb is a consequence of the non-monotonous rotation of the Stokes vector about the time-dependent magnetic field \eqref{B_eff}. Interestingly, in contrast to the LC regime, the relative strength of the spectral components are not fixed but depend on the trajectory on which the dimer resides.  
 The spectra corresponding to the adjacent periodic orbits are shown in Fig.~\ref{fig4}a.
Close to stationary solutions, there is one considerable peak originating either from $\rm{FP}_1$ (the red curve) or $\rm{FP}_2$ (the blue curve). Away from the FPs, the spectrum has two strongest peaks and several equidistant multiplets. The intensity accumulated in the multiplets increases in the strong pump regime due to the contribution of the parametric processes triggered by the conservative non-linearity. When switching the dimer between two FP solutions, the positions of the strongest peaks remain fixed but their relative weights vary.

The weights of the lines corresponding to FPs can be defined as the total intensity $\left|S_{1,2}(\omega) \right|^2 = \left|\int \psi_{1,2}(t) e^{-i\omega t} dt\right|^2$ concentrated in the spectral line. The weights extracted from the numerical simulations are shown in Fig.~\ref{fig4}b. 
Since the $\tilde{\mathcal{P}}\mathcal{T}$-operator breaks the inversion symmetry between the modes, one has to distinguish the spectra $S_1(\omega)$  and  $S_2(\omega)$ emitted by different modes although the corresponding dynamics are qualitatively similar. Fig.~\ref{fig4}b demonstrates that the occupation of the stationary solutions can be tuned continuously at the fixed system parameters. In practice, it can be done with the coherent pulse driving the system at the frequency of the stationary solution. The numerical simulations of this effect are presented in the Appendix~\ref{AppSwitching}.

Note that the oscillating solutions  disappear if the conditions \eqref{Eq_detuning} and \eqref{PH_conditions} are violated. In particular, it happens if the pseudo-Hermiticity of $\hat{\mathcal{H}}_{\rm PH}$ is spontaneously broken at $|\gamma|>|J|$, see Fig.~\ref{fig4}c. The multi-component spectrum (with two dominating linear eigenstates) exists below the symmetry breaking threshold, $\gamma\leq J$. However at $\gamma>J$, the dimer resides to the single stable FP.

\section{Discussion}

\begin{figure}
    \includegraphics[width=\linewidth]{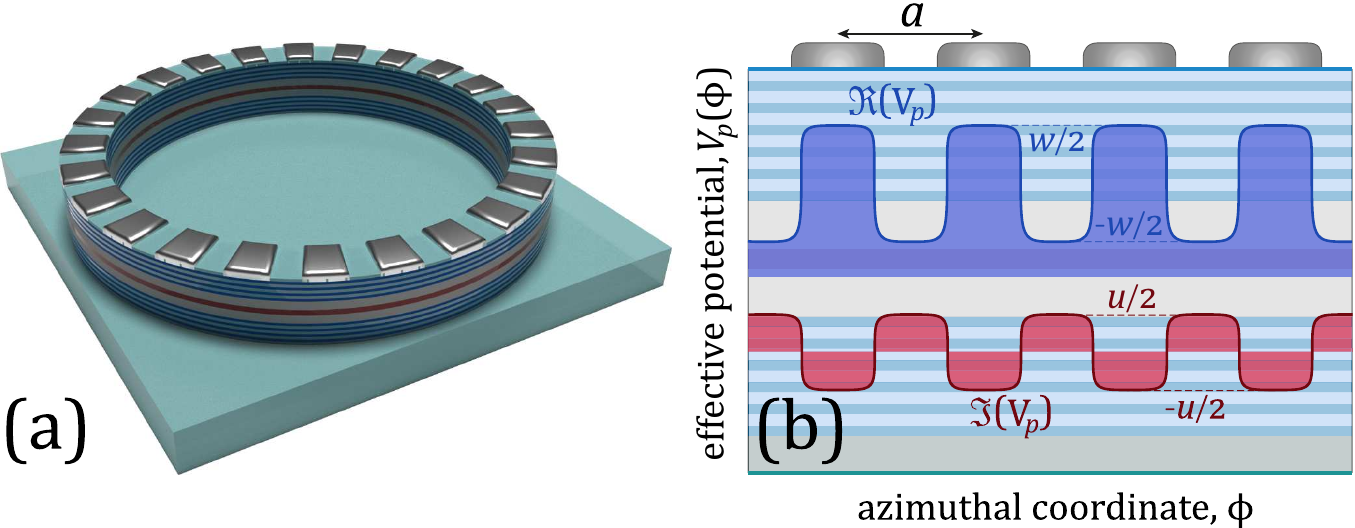}
    \caption{(a) The sketch of a ring-shaped microcavity capable of supporting of the pseudo-Hermitian dynamics of the polariton condensate. (b) Spatial distribution of the ridge-shaped complex potential $V_p(\mathbf{r})$ originating from the metallic fringes on the top. Similar potentials can be also induced optically with use of a spatio-optical modulator. }
    \label{Fig:scheme}
\end{figure}

In this section we discuss possible experimental realizations of the predicted pseudo-conservative dynamics. Several conditions imposed onto both linear and nonlinear coupling and detuning of dimer modes narrows down the range of available systems where the effect could be observed.
The condition on the anti-Hermitian nonlinear operator form \eqref{Eq_nonl_oper} is satisfied in cavities with periodic complex potentials that can be induced by metallic fringes on the top of the cavity \cite{Nalitov2017}. Alternatively, a periodic complex potential  can be created by a spatially inhomogeneous optical pumping achieved with use of spatio-light modulator \cite{Berloff2017}.
The corresponding two-mode basis in this case is the pair of counter-propagating plane waves with wave vectors $\pm\pi/a$ with $a$ being the potential period, see Fig.~\ref{Fig:scheme}.
However, the condition on the linear coupling \eqref{Eq_detuning} is out of reach as the dimer detuning $\varepsilon$ is absent due to the symmetry of the system.
On the other hand, this symmetry can be externally broken, moreover, the real and imaginary parts $\varepsilon$ and $\gamma$ can be controlled independently.

The asymmetry $\gamma$ of the effective pumping can be induced by pseudodrag effect \cite{Chestnov2019} caused by motion of exciton reservoir, pumping the condensate through bosonic stimulated scattering.
This motion, in turn, can be imparted either by electric current in proximity due to Coulomb forces or by incoherent optical pumping of excitons.
In the vicinity of the condensation threshold for one of the doublet states the splitting can be estimated as $\gamma = 4 \pi \gamma_c \lambda_\text{th}^2 k_r/a$, where $\lambda_\text{th}$ is the exciton reservoir de Broglie wavelength, $k_r$ is its mean wavevector value and $\gamma_c$ is the cavity loss-rate.

The real energy splitting $\varepsilon$ can be induced by the motion of the potential acting upon the condensate of polaritons. The rotation of the optically induced potential in the case of a finite size ring configuration, see Fig.~\ref{Fig:scheme}a, would be one way to break the symmetry between clock-wise and anti-clockwise current solutions.
Alternatively, stirring potential similar to those used for rotating excitons \cite{Hasling2015}, can be imposed by periodic alternation of voltage at the metallic overlayer fringes.
Regardless of realization, the energy splitting can be estimated as $\varepsilon = 2\pi \hbar v/a$, where $v$ is the potential shift velocity.

Finally, the coupling matrix elements $J=w/\pi$ and $\varkappa = u/\pi$ are controlled by the amplitudes of real and imaginary parts of the periodic effective potential $V_p(\mathbf{r})$, see Fig.~\ref{Fig:scheme}b.
The linear condition \eqref{Eq_detuning} then transforms into the condition on the reservoir and the potential motion characteristics:
\begin{equation}
     k_r v = -{a^2 u w \over 8 \pi^4 \hbar \gamma_c \lambda_\text{th}^2}.
\end{equation}

\section{Conclusions}
The $\mathcal{PT}$-symmetry endows open systems  with the dynamics which can be naturally expected for their isolated counterparts. However, the most important practical applications of this phenomenon take advantage of the symmetry-broken phase to the detriment of the pure $\mathcal{PT}$-symmetric regime \cite{Chang2014,Ramezani2010,Lin2011,Guo2009,Feng2014,Peng2014,Hodaei2014}.
Our findings demonstrate that the dynamics typical of the isolated systems can be efficiently reproduced in a driven-dissipative system composed by two quantum states characterized by different gain (or loss) rates. This implies a combination of the passive $\mathcal{PT}$-symmetry with the specifically engineered mechanism of the gain saturation. The behaviour of the normalized state vectors possesses all the phenomenology of the conservative dynamics. In contrast to the conventional $\mathcal{PT}$-symmetry, the discussed phenomena can be realized at an arbitrary  level of pumping of a driven dissipative system, such as a polariton laser above the threshold.

\begin{acknowledgments}
This work is supported by the Westlake University (Project No.~041020100118) and from Program 2018R01002 funded by the Leading Innovative and Entrepreneur Team
Introduction Program of Zhejiang. I.C. acknowledges funding from National Natural Science Foundation of China (Grant No. 12050410250). The support from
RFBR grant 21-52-10005, from the Grant of the President of
the Russian Federation for state support of young Russian scientists
No. MK-5318.2021.1.2 and from the state task  in the scientific activity
project 0635-2020-0013 is acknowledged. 
Y.G.R. acknowledges the support from CONACYT (Mexico) Grant No.\ 252808 and from PAPIIT-UNAM Grant No.\ IN 106320.
A.K. acknowledges the support from the Road Map for Quantum Computing program of the Rosatom. A.N. acknowledges the support from the European Union’s Horizon 2020 research and innovation
programme under the Marie Sk\l{}odowska-Curie grant agreement No. 846353. %
\end{acknowledgments}

\appendix

\section{Two-state ansatz}\label{AppA}

Here we describe how the spinor equation \eqref{Pol_EQ} can be derived from the two-dimensional model of the trapped exciton-polariton condensate. We use the driven-dissipative model \cite{Wouters2007} which describes the mean-field dynamics by the Ginzburg-Landau equation for the order parameter $\Psi(\mathbf{r})$ coupled to the kinetic equation for the density   $n(\mathbf{r})$ of incoherent excitons:
\begin{subequations}\label{GPE}
	\begin{eqnarray}
		i\partial_t\Psi  &=& \left(-\frac{1}{2m}\bm{\nabla}^2 + V_p(\mathbf{r}) + g_r n\right)\Psi + \notag\\
		&+& \frac{i}{2}(Rn-\gamma_c)\Psi + g_c \left|\Psi\right|^2\Psi, \label{GPE_Psi} \\
		 \partial_t n &=& P(\mathbf{r}) - (\gamma_r + R\left|\Psi\right|^2)n.
	\end{eqnarray}
\end{subequations}
Here $\hbar=1$, $m$ is the polariton effective mass, $g_c$ and $g_r$ describe the interaction of polaritons between themselves and with incoherent excitons, respectively, $R$ is the rate of scattering from the reservoir to the condensate, $\gamma_c$ characterizes the cavity losses and $\gamma_r$ stands for the reservoir relaxation rate.

The equation (\ref{GPE}b) is excluded by assuming that the reservoir relaxation $\gamma_r$ is fast. This approximation corresponds to the conditions of the recent experiments  \cite{Kim2020}. In this case, the reservoir density follows instantly the variation of the condensate distribution such as $n(\mathbf{r}) \approx P(\mathbf{r})/\gamma_r - P(\mathbf{r}) R |\Psi(\mathbf{r})|^2/\gamma_r^2$. When substituted in Eq.~\eqref{GPE_Psi}, this expression yields the nonlinear dissipative term which corresponds to the relaxation rate governed by the condensate density. 

Then we use the two-level ansatz for the condensate order parameter:
\begin{equation}\label{ansatz}
	\Psi(\mathbf{r}) = \psi_1(t)\varphi_1(\mathbf{r}) + \psi_2(t)\varphi_2(\mathbf{r}),
\end{equation}
where the mode distribution functions obey the normalization conditions $\int |\varphi_{1,2}|^2d\mathbf{r}=1$. We substitute the ansatz \eqref{ansatz} into Eq.~\eqref{GPE_Psi} and integrate out spatial degrees of freedom assuming that the overlapping integrals $\int \varphi_i \varphi_j^* d\mathbf{r}$ are small compared to unity which implies that the modes are nearly orthogonal. This procedure yields two coupled equations for the mode amplitudes:
\begin{eqnarray}\label{Eq_coupled_osc}
\notag		i\partial_t \psi_{1,2} &=& \left(\varepsilon_{1,2} + i/2\left( \mathcal{W}_{1,2} - \gamma_c\right) \right) \psi_{1,2} + \left(J + i\varkappa\right)\psi_{2,1} \\
&+& \chi\left(\alpha_{1,2} |\psi_{1,2}|^2 + |\psi_{2,1}|^2 \right)\psi_{1,2} \\
\notag 		&-& i\nu \left(\beta_{1,2} |\psi_{1,2}|^2 + |\psi_{2,1}|^2 \right)\psi_{1,2},
\end{eqnarray}
where
\begin{eqnarray}
	\notag 	\varepsilon_{1,2}&=&-\frac{1}{2m} \int \varphi_{1,2}^*\nabla^2\varphi_{1,2} d\mathbf{r} + \int \varphi_{1,2}^* V(\mathbf{r})\varphi_{1,2} d\mathbf{r}, \\
	\notag V(\mathbf{r})&=&V_p(\mathbf{r}) + \frac{g_r P(\mathbf{r})}{\gamma_r}, \\
	\notag J+i\varkappa &=& -\frac{1}{2m} \int \varphi_1^*\nabla^2\varphi_2 d\mathbf{r} + \int \varphi_1^* V(\mathbf{r})\varphi_2 d\mathbf{r}, \\
		\mathcal{W}_{1,2} &=& \frac{R}{\gamma_r} \int P(\mathbf{r}) |\varphi_{1,2}(\mathbf{r})|^2 d\mathbf{r}, \\
	\notag \nu &=& \frac{R^2}{\gamma_r^2} \int P(\mathbf{r}) |\varphi_1(\mathbf{r})|^2|\varphi_2(\mathbf{r})|^2 d\mathbf{r}, \\
	\notag \chi &=& \int \left(g_c - {P(\mathbf{r})R g_r}/{\gamma_r^2}\right)|\varphi_1(\mathbf{r})|^2|\varphi_2(\mathbf{r})|^2 d\mathbf{r}, \\
	\notag \alpha_{1,2} &=& \frac{\int \left(g_c - {P(\mathbf{r})R g_r}/{\gamma_r^2}\right) |\varphi_{1,2}(\mathbf{r})|^4 d\mathbf{r}}{\int \left(g_c - {P(\mathbf{r})R g_r}/{\gamma_r^2}\right) |\varphi_{1}(\mathbf{r})|^2 |\varphi_{2}(\mathbf{r})|^2 d\mathbf{r}},
	 \\
	 \notag \beta_{1,2} &=& \frac{\int P(\mathbf{r}) |\varphi_{1,2}(\mathbf{r})|^4 d\mathbf{r}}{\int P(\mathbf{r}) |\varphi_{1}(\mathbf{r})|^2 |\varphi_{2}(\mathbf{r})|^2 d\mathbf{r}}. 
\end{eqnarray}

Defining the gain asymmetry parameter $\gamma = (\mathcal{W}_1-\mathcal{W}_2)/4$, the net gain  $\mathcal{W} = (\mathcal{W}_1+\mathcal{W}_2)/4 - \gamma_c/2 $ and the energy detuning  $\varepsilon = (\varepsilon_{1} - \varepsilon_{2})/2$, we recast Eq.~\eqref{Eq_coupled_osc} in the form of the spinor equation \eqref{Pol_EQ} with the Hamiltonian~\eqref{H_pol}.

\section{Pseudo-conservative dynamics in the presence of the time-dependent reservoir}\label{AppB}

So far, we assumed the presence of the gain asymmetry between the uncoupled modes which is required to fulfill the $\mathcal{PT}$-symmetry criteria when the coupling parameter is complex, see Eq.~\eqref{Eq_detuning}. However, the pseudo-conservative dynamics can  also occur in a dimer with identical gain parameters $\mathcal{W}_1=\mathcal{W}_2=\mathcal{W}_p$ provided that the loss rates are different, $\gamma_{c1} \neq \gamma_{c2}$. The required asymmetry of the loss rates can be achieved with the use of various approaches for the spatial loss-engineering \cite{Stroev2020}. A replacement of the unequal individual gains with the unequal losses has no effect on the pseudo-Hermitian dynamics predicted by Eq.~\eqref{Pol_EQ}. The Hamiltonian \eqref{H_pol} retains its form with the new definitions for the asymmetry $\gamma= (\gamma_{c2} - \gamma_{c1})/4$ and the net gain $\mathcal{W} = \mathcal{W}_p/2 - (\gamma_{c1} + \gamma_{c2})/4$ parameters.

The key assumption made in the main text is that the gain saturation occurs with no retardation. For the trapped polariton condensate, it implies that the reservoir instantly adjusts to the changes of the condensate density. It is instructive to demonstrate that the retardation of the gain saturation mechanism can have no impact on the existence of the continuum of periodic solutions.

Here we account for the influence of the reservoir dynamics on the parameter range when adiabatic elimination is impossible. The gain-saturation effect is described by the term $Rn|\Psi|^2$ (see Eq.~(\ref{GPE}b)) which accounts for the scattering from the reservoir to the condensate. Substituting $n(\mathbf{r},t) = N(\mathbf{r},t)n_r(t)$, using the two-mode ansatz for the condensate density and assuming the loss rates asymmetry between the uncoupled modes,  we obtain the model which explicitly accounts for the time-dependence of the reservoir population:
\begin{subequations}\label{Eq_res_model}
    \begin{eqnarray}
        i \partial_t \bm{\psi} &=&  \hat{\mathcal{H}}_{\rm PH} \bm{\psi} +i \mathcal{W}  \sigma_0 \bm{\psi} + g n_r \bm{\psi}, \\ 
        \partial_t n_r &=& P_0 - \gamma_r n_r - \bar{R} n_r \left\| \bm{\psi} \right\|  ,
    \end{eqnarray}
\end{subequations}
where $P_0=\int P(\mathbf{r}) d\mathbf{r}$, the net gain is governed by the reservoir density $\mathcal{W} \equiv \mathcal{W}(n) = \bar{R} n_r(t)/2 - (\gamma_{c1} + \gamma_{c2})/4$. Here we assumed that $\int N(\mathbf{r}) |\varphi_1(\mathbf{r})|^2 d\mathbf{r} = \int N(\mathbf{r}) |\varphi_2(\mathbf{r})|^2 d\mathbf{r} = Q$. The value of $Q$ governs the effective rate  of the exciton scattering to the condensate  $\bar{R}=RQ$. The last term in Eq.~(\ref{Eq_res_model}a) stems from the condensate interaction with the hot excitons from the reservoir. The effective  strength of this interaction is $g=g_r Q$.

Equation~(\ref{Eq_res_model}a) can be transformed to the Stokes vector basis. The result is equivalent to Eq.~\eqref{Eq_pseudospin}   
but with $\mathbf{B} = 2\left(J,0,\delta\varepsilon\right)^\intercal$ and $\mathbf{E} = 2\left(\varkappa,0,\gamma\right)^\intercal$. The reservoir entering the equation in the effective gain parameter $\Gamma=2\mathcal{W}=\bar{R}n_r(t)-(\gamma_{c1} + \gamma_{c2})/2$ only. However, the evolution of the normalized Stokes vector $\mathbf{s} $ does not depend on $\Gamma$ and thus on the reservoir dynamics as well. Indeed, even in the presence of the time-dependent reservoir, the behaviour of $\mathbf{s}$ is governed by Eq.~\eqref{Eq_eff}, where the effective field ${\mathbf{b}}$ is independent of $\mathcal{W}$. Note that this result is valid at any reservoir relaxation rate $\gamma_r$.

\section{Switching between the non-isolated orbits by the coherent resonant pulses}\label{AppSwitching}

\begin{figure}
    \includegraphics[width=\linewidth]{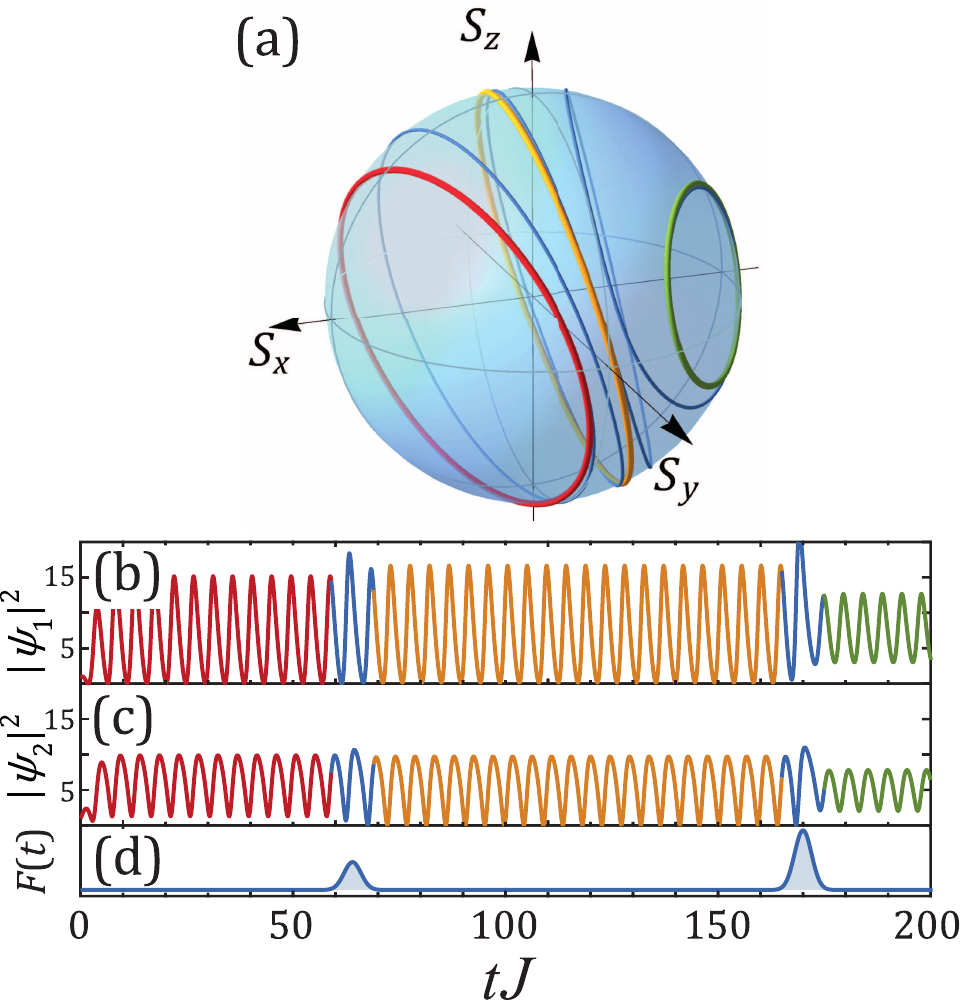}
    \caption{Switching between the neutrally stable pseudospin trajectories. The dynamics is governed by Eqs.~\eqref{Eq_res_model} which accounts for the presence of the reservoir. (a) -- the  pseudospin projections on the Bloch sphere with a unitary radius. Red, orange and green circles correspond to the stable rotation of the pseudospin while the blue lines indicate the trajectories  when the external coherent field is switched on. The corresponding mode populations $|\psi_1|^2$ and $|\psi_2|^2$ are shown on (b) and (c), respectively. Two coherent pulses $F(t)$ illustrated in the panel (d) drive mode $\psi_1$ at $t=64$ and $t=170$. Time is measured in units of the inverse Josephson coupling strength $|J|^{-1}$. The amplitudes of the first and the second pulses are $F_0 = 0.35J$ and $F_0 = 0.75J$, respectively. The duration of both pulses is $\tau=2J^{-1}$. Other parameters of the model Eq.~\eqref{Eq_res_model} are $P=250$, $\gamma_r=20$, $\varkappa=0.2$, $J=-1$, $\varepsilon=-\varkappa J/\gamma$, $\mathcal{W}_1=2.5$ and $\mathcal{W}_2=5.5$, where all energies are measured in units of $|J|$.}
    \label{FigS1}
\end{figure}

The presence of the continuum of periodic orbits can be also demonstrated by introducing the coherent pumping pulse $F(t)$ which drives a single mode triggering switching between stable trajectories. In particular, we take it as a Gaussian pulse of the duration $\tau$, which drives the mode $\psi_1$ at the time moment $t_0$, $F(t) = F_0 \exp\left( - (t - t_0)^2\left/2 \tau^2\right.\right)$. 
The relevant dynamics is demonstrated in Fig.~\ref{FigS1}.

\section{The phase diagram at negative Josephson-to-dissipative coupling ratio}\label{AppNegJtDratio}

\begin{figure}
    \includegraphics[width=\linewidth]{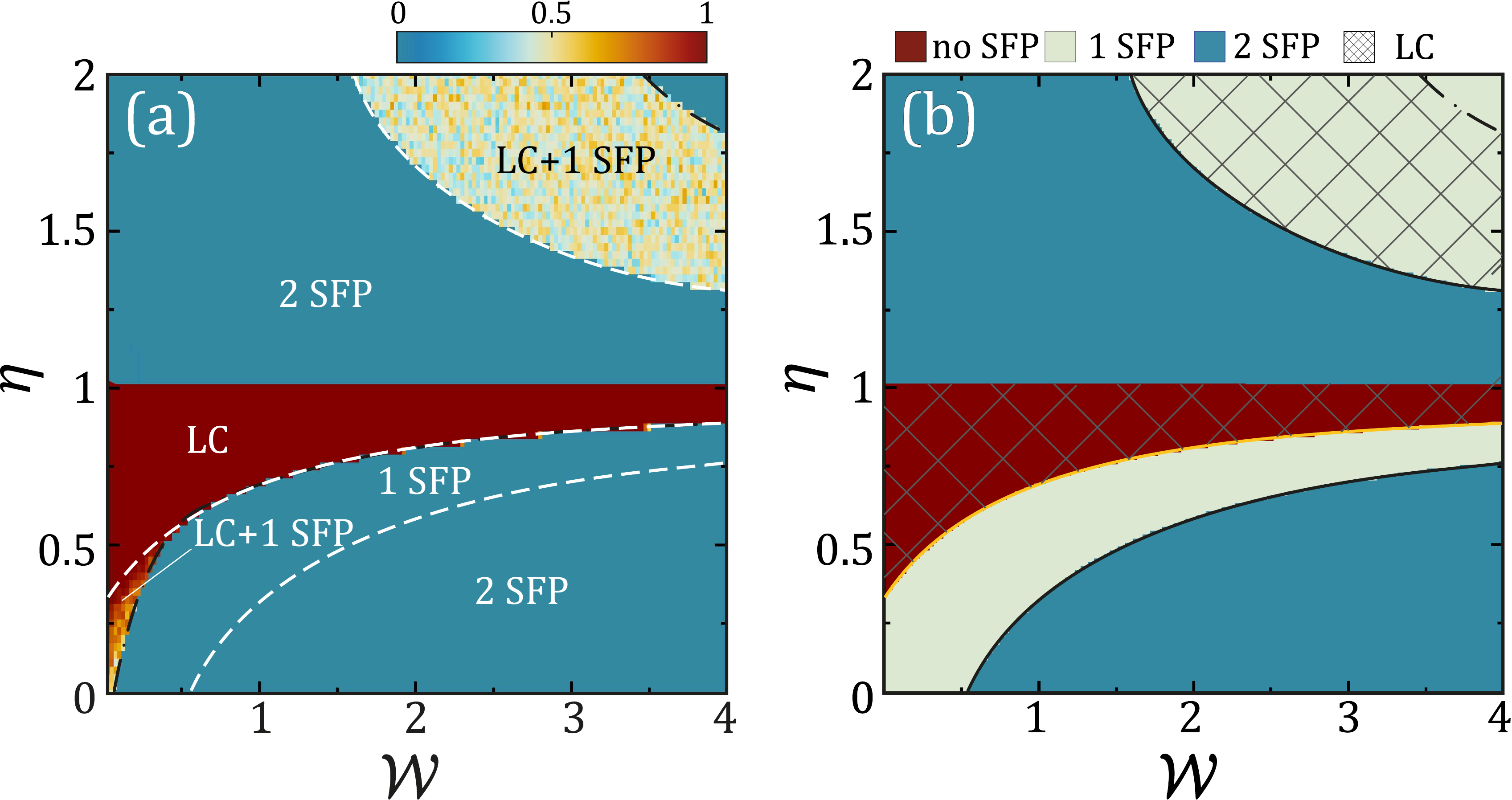}
    \caption{(a) A phase diagram illustrating the limit cycle formation probability. All the parameters  are the same as in Fig.~\ref{Fig:PD} but $\varkappa=-0.1$. The dimer is excited starting from the noisy initial conditions. The data were averaged over 50 realizations for each point. (b) The chart of attractors on the $(\eta,\mathcal{W})$-plane.}
    \label{Fig:PD_negative}
\end{figure}

Figure \ref{Fig:PD} demonstrates a typical example of the phase diagram. It illustrates different types of bifurcations which occur in the  $\tilde{\mathcal{P}}\mathcal{T}$-symmetric system with saturated gain. The used parameter space is chosen to highlight the impact of the nonlinear effects. The positions of the domain boundaries depend on the parameters of the linear Hamiltonian \eqref{H_pol} as well. In particular, the abrupt changes occur if one inverts the sign of the Josephson-to-dissipative coupling ratio. An example of the phase diagram for this case is shown in Fig.~\ref{Fig:PD_negative}.  As in the main text, we assume that the $\tilde{\mathcal{P}}\mathcal{T}$-symmetry of $\hat{\mathcal{H}}_{\rm PH}$ holds.
 
We keep following the convention that the periodic trajectories which encircle the trivial FPs are called LCs of the first type. The LC of the second type has the trajectory which passes around the symmetry-breaking FPs. 
When the Josephson $J$ and the dissipative $\varkappa$ couplings have different signs, the LC of the first type occurs at $\eta<1$ in contrast to the case $J/\varkappa>0$ shown in the main text. However the location of the LC of the second type remains unaffected if the sign of the relation $J/\varkappa$ is inverted. 
	
\section{Nonlinear dynamics of the $\tilde{\mathcal{P}}\mathcal{T}$-symmetric dimer}\label{AppPhaseDiagramDetails}	
	
\begin{figure}
    \includegraphics[width=\linewidth]{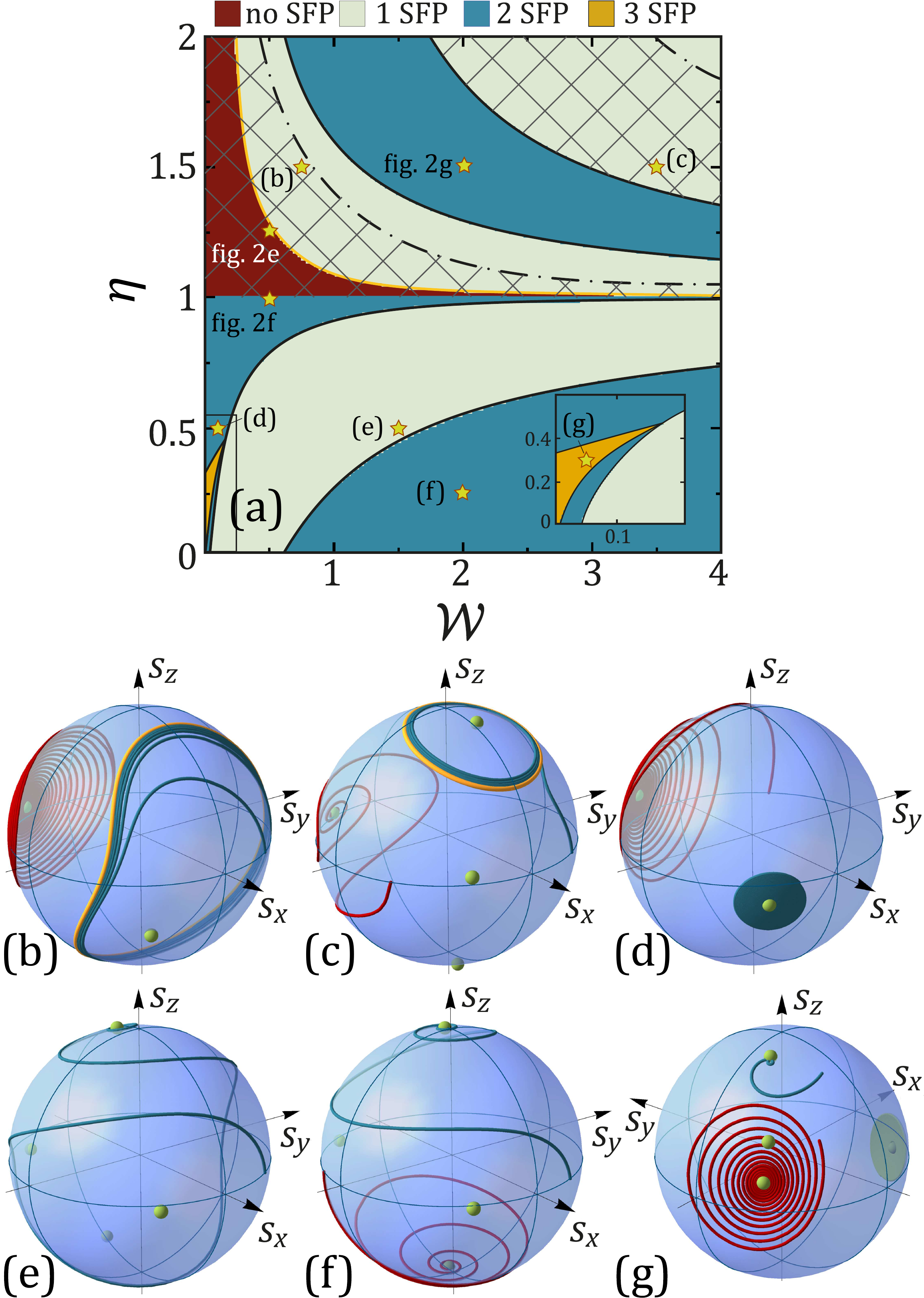}
    \caption{Nonlinear dynamics of the $\tilde{\mathcal{P}}\mathcal{T}$-symmetric dimer. (a) The chart of attractors on the $(\eta,\mathcal{W})$-plane for the same parameters as in Fig.~\ref{Fig:PD}b. (b)-(g) The scenario of approaching stable attractors in different dynamical regimes indicated by the yellow stars on (a). The green dots correspond to the FP solutions. The orange orbits indicate  the trajectories of the stable LCs. }
    \label{Fig:PD_detalied}
\end{figure}

Here we give an extended description of the nonlinear dynamics of the $\tilde{\mathcal{P}}\mathcal{T}$-symmetric dimer discussed in Sec.~\ref{Sec:IV}. For the sake of clarity, we focus on the case of positive ratio of the Josephson-to-dissipative coupling. Figure~\ref{Fig:PD_detalied}a demonstrates the very same chart of attractors on the $(\eta,\mathcal{W})$-plane which is shown in Fig.~\ref{Fig:PD}b but with the examples of the Stokes vector evolution in every domain. The domains on the chart are distinguished by the type and the number of stable attractors. The number of FPs which are stable against small perturbations varies from zero to three. Besides, there are  two types of stable LCs whose existence domains do not overlap for the considered parameters. However, as the strength  of dissipative coupling $\varkappa$ approaches $J$, the boundary of the LC of the first type shifts towards larger values of $\mathcal{W}$. In this case both LCs can coexist at the same parameters. In such multistability domains,  formation of the particular state depends on whether the initial position of the Stokes vector resides in its basin of attraction or not.

Three scenario of the dimer evolution were illustrated in Figs.~\ref{Fig:PD}e-f. The most important one corresponds to the rotation on a circular trajectory which belongs to the continuous band of closed orbits in the pseudo-conservative regime, see Fig.~\ref{Fig:PD}f. This regime is destroyed at any $\eta \neq 1$. In particular, above $\eta=1$ the Stokes vector slowly approaches a single stable LC (of the first kind) which is a final state regardless of the initial pseudo-spin position, see Fig.~\ref{Fig:PD}e. With the further increase of $\eta$ a subcritical Hopf bifurcation (the yellow line in Fig.~\ref{Fig:PD_detalied} between the vinous and the grey domains) changes stability of one of the trivial FP solutions and gives birth to the unstable LC. In this case, the LC of the first type coexists with the nodal point, see Fig.~\ref{Fig:PD_detalied}b. The unstable LC then  moves towards the stable one as either $\mathcal{W}$ of $\eta$ increase. When they collide (the dash-dotted in Fig.~\ref{Fig:PD_detalied}), the stable LC solution disappear while the trivial FP remains the only stable attractor.

At strong gain $\mathcal{W}$, a couple of symmetry-broken FPs appear in the saddle-node bifurcation. It increases the total number of stable attractors by one. Reaching either of two stable FP solutions in this regime is shown in  see Fig.~\ref{Fig:PD}g. With the further increase of the gain, the stable symmetry-broken state changes to an unstable spiral surrounded by a small isolated closed orbit. This is the LC of the second type, whose existence domain is indicated by the hatched area in the upper-right corner of Fig.~\ref{Fig:PD_detalied}a. The relevant dynamics of the Stokes vector is illustrated in Fig.~\ref{Fig:PD_detalied}c. The dimer either resides on the LC or approaches the nodal point depending on its initial position. The orbit of the LC expands as the gain increases further. At large $\mathcal{W}$, it touches the saddle symmetry-broken FP and disappear in the homoclinic bifurcation leaving the trivial nodal point a sole stable solution. 

Our numerical analisys didn't reveal the presence of the stable LC solutions below $\eta=1$. Instead there is a plethora of stable FPs. In particular, in the vicinity of $\eta=1$, there are two trivial FPs which stable simultaneously, see Fig.~\ref{Fig:PD_detalied}d. Away from $\eta=1$, one of the trivial FP losses its stability giving birth to the large domain where a single  nodal point exists, see Fig.~\ref{Fig:PD_detalied}e. An appearence of the symmetry broken states brings another stable attractor, see the domain in the lower-right corner of Fig.~\ref{Fig:PD_detalied}a and an example of the corresponding dynamics in Fig.~\ref{Fig:PD_detalied}f. The most exotic case is realized when the symmetry-broken states appear at high imbalance of the self- and cross-phase modulation coefficients and at weak gain (the orange domain in the inset of Fig.~\ref{Fig:PD_detalied}a). In this case both trivial and a single symmetry broken solutions are stable, see Fig.~\ref{Fig:PD_detalied}g.


%

\end{document}